\documentclass[12pt,preprint]{aastex}
\usepackage{natbib}
\usepackage{amsmath}                
\usepackage{amsfonts}               
\usepackage{amssymb}                
\usepackage{multirow}
\usepackage[usenames,dvipsnames]{color}

\newcommand{\href}[2]{#2}
\def\astrobj#1{#1}
\def\kms{km\,s${}^{-1}$}
\def \s{~\rm{s}}
\def \g{~\rm{g}}
\def \yr{~\rm{yr}}
\def \cm{~\rm{cm}}
\def \km{~\rm{km}}
\def \kpc{~\rm{kpc}}
\def \Mpc{~\rm{Mpc}}
\def \erg{~\rm{erg}}
\newcommand{\mbh}{M_{\rm BH}}
\newcommand{\mgal}{M_{\rm G}}
\newcommand{\rst}{R_\textrm{strip}}
\newcommand{\rhog}{\rho_\textrm{g}}
\newcommand{\rhor}{\rho_\textrm{r}}

\begin{document}

\title{The influence of mergers and ram-pressure stripping\\ on black hole-bulge correlations}
\shorttitle{Mergers and ram-pressure stripping}

\author{Yonadav Barry Ginat\altaffilmark{1}, Yohai Meiron\altaffilmark{2,3,4} and Noam Soker\altaffilmark{1}}
\shortauthors{Ginat, Y. B., Meiron, Y., and Soker, N.}

\affil{${}^1$Department of Physics, Technion -- Israel Institute of Technology, Haifa, 3200003, Israel; ginat@campus.technion.ac.il; soker@physics.technion.ac.il\\
${}^2$Institute of Physics, E\"otv\"os University, P\'azm\'any P. s. 1/A, Budapest, 1117, Hungary\\
${}^3$Kavli Institute for Astronomy and Astrophysics at Peking University, 5 Yiheyuan Rd., Haidian District, 100871, Beijing, China\\
${}^4$National Astronomical Observatories of China, Chinese Academy of Sciences, 20A Datun Rd., Chaoyang District, 100012, Beijing, China
}

\begin{abstract}
We analyse the scatter in the correlation between super-massive black hole (SMBH) mass and bulge stellar mass of the host galaxy, and infer that it cannot be accounted for by mergers alone.
The merger-only scenario, where small galaxies merge to establish a proportionality relation between the SMBH and bulge masses, leads to a scatter around the linear proportionality line that increases with the square root of the SMBH (or bulge) mass.
By examining a sample of {{{{103}}}} galaxies we find that the intrinsic scatter increases more rapidly than expected from the merger-only scenario. The correlation between SMBH masses and their host galaxy properties is therefore more likely to be determined by a negative feedback mechanism that is driven by an active galactic nucleus.
We find a hint that some galaxies with missing stellar mass reside close to the centre of clusters {{{{and speculate}}}} that ram-pressure stripping of gas off the young galaxy as it moves near the cluster centre, might explain the missing stellar mass at later times.
\end{abstract}


\section{INTRODUCTION}
\label{sec:intro}

The correlations of super-massive black hole (SMBH) masses with properties of their host galaxies are a hot research topic (see, e.g., reviews by \citealt{Graham2015, KormendyHo2013}). For these correlations the stellar mass (or luminosity) of the spheroidal component (hereafter referred to as the bulge), $M_{\rm G}$ and the stellar velocity dispersion of the bulge, $\sigma$ are mostly used (e.g. \citealt{KormendyRichstone1995,  Laor2001, Magorrian_etal1998, Hu2009, Graham2009, McConnellMa2013, Aversaetal2015, Savorgnanetal2016} for $\mgal$; and \citealt{Gebhardt_etal2000, MerrittFerrarese2001, Hu2008, Graham2008a, Graham2008b,   Gultekin_etal2009, Shen_etal2008, GrahamScott2013, Saxtonetal2014, SavorgnanGraham2015a, Shankaretal2016} for $\sigma$). Correlations with other properties are also studied (e.g., \citealt{FeoliMancini2011,Sagliaetal2016}), such as the number of globular clusters (e.g.,  \citealt{BurkertTremaine2010}), the kinetic energy parameter $M_{\rm G} \sigma^2$ (e.g., \citealt{Feoli_etal2010, ManciniFeoli2012, Benedettoetal2013, Feoli2014}), and the momentum parameter $\mu = M_{\rm G} \sigma/c$ (e.g., \citealt{SokerMeiron2011, Lahavetal2011, Cen2012}). {{{{Most recently, \cite{Sagliaetal2016} found the best correlation to be between the SMBH mass and $M^{1/2}_{\rm G} \sigma^2$, as also discussed by \cite{Hopkinsetal2007a, Hopkinsetal2007b}.}}}} The correlations are chiefly assumed to take the form of power laws -- linear when plotted on a log-log scale.

Some studies argued that the $M_{\rm BH}$--$M_{\rm G}$ correlation could be brought about by mergers between many small galaxies (e.g., \citealt{Peng2007, Jahnke2011, Gaskell2011}), and that a feedback process based on active galactic nuclei (AGN; which seems to affect both the stars' evolution and that of the SMBH, \citealt{Aravetal2015}) is superfluous (e.g., \citealt{Peng2007}). {{{{Mergers may play a more modest role: \cite{Sagliaetal2016} for example, argue that mergers may be important only in explaining the displacement of core ellipticals from the correlations.}}}}

It is expected (see Section \ref{sec:scatternew}) that if only galaxy mergers lead to significant SMBH growth, the intrinsic scatter of BH masses would increase as $n^{1/2}$, where $n$ is the number small galaxies, the `building blocks', that merged to form the final galaxy. As the mass increases in proportion with $n$, the relative scatter in linear scale should decrease as $n^{-1/2}$ in the merger-only scenario.

Drawing upon these arguments on intrinsic scatter evolution, \cite{Lahavetal2011} conducted a preliminary study of the merger scenario for BH-mass-bulge-mass correlation. They examined a sample of 86 galaxies and found that the intrinsic scatter increases with mass more rapidly than expected from the merger-only scenario. Hence, they concluded, the merger-alone scenario cannot account for the $M_{\rm BH}$--$M_{\rm G}$ correlation.
Several other studies followed, and strengthened the conclusion of \cite{Lahavetal2011} on the limited role mergers play in forming the correlations (e.g., \citealt{Shankaretal2012, Debattistaetal2013, Aversaetal2015, SavorgnanGraham2015a, Mechtleyetal2016}), and more generally in building galaxies (e.g., \citealt{Narayananetal2015}; cf. \citealt{Shankaretal2009} for BH growth below $10^9 M_\odot$).

In the present study we update the analysis of \cite{Lahavetal2011} of the scatter in the $M_{\rm BH}$--$M_{\rm G}$ correlation, and go further to discuss how the correlation is affected by inflation of gas due to AGN activity. We describe the sample of galaxies we use in Section \ref{sec:samplenew}, and in Section \ref{sec:scatternew} we analyse the implications to the merger scenario. In Section \ref{sec:outliers} we consider a certain class of galaxies -- those residing near cluster centres -- and their $\mbh$-$\mgal$ relation relative to the general one. We present our summary and conclusions in Section \ref{sec:summary}.

\section{THE SAMPLE}
\label{sec:samplenew}

The sample of galaxies is assembled from the following sources: first is the list of galaxies compiled by \cite{SavorgnanGraham2015a, SavorgnanGraham2015b, Savorgnanetal2016} (based on \citealt{GrahamScott2013}), \cite{KormendyHo2013} and \cite{McConnellMa2013}, (SG, KH and MM respectively) and other galaxies taken from individual studies (see below). The data appear in Table \ref{tab:Galaxytabel}. All {{{{(106)}}}} galaxies have measured $\sigma$, and {{{{103}}}} galaxies have measured bulge mass, which is the entire stellar mass in elliptical galaxies. All galaxies in \cite{McConnellMa2013} are included in \cite{KormendyHo2013}, as well as some galaxies in the sample of \cite{SavorgnanGraham2015a, SavorgnanGraham2015b, Savorgnanetal2016} (which were combined for values of both $\sigma$ and $M_G$); however, the sources differ regarding some values for the same parameters. In such cases averages were calculated and used for the plots given here. As \cite{McConnellMa2013} do not give error estimates for $M_{\rm G}$, the errors were taken from \cite{KormendyHo2013} and, if present, from \cite{SavorgnanGraham2015a, SavorgnanGraham2015b}. Errors for the average values of the bulge mass and $\sigma$ from $p$ sources were estimated according to
\begin{equation}
    \delta x = \frac{1}{p}\sqrt{ \sum_{i=1}^p (\delta x_i)^2} ,
      \label{eq:deltax}
\end{equation}
where $x$ stands for the variable whose errors are to be calculated. In the few cases where the standard error of the mean were bigger, it was used as the error. The errors for the black hole masses are the average values of the higher and lower estimates.

{{{{ In this work we neglect the possible effects of non-Gaussianity of the errors and their cross-correlation, as a thorough discussion of these is beyond the scope of this work. We note that recently \cite{Sagliaetal2016} calculated the error covariance matrix of various quantities and found that the off-diagonal terms were smaller than the diagonal terms, adding that ignoring them is not a bad approximation. }}}}

There are several exceptions. Two individual BHs that do not appear in any of the three compilations; these are the BHs in NGC 1271 which we take from \citet{Walshetal2015a} {{{{and its bulge mass from \cite{Grahametal2016a}; and last, the BH in M60-UCD1 \citep{Sethetal2014}. Besides these two, the values adopted for the mass of the black hole and spheroid of NGC 1277 are the newest estimates presented by \cite{Grahametal2016}, although a higher BH mass was suggested by \cite{Walshetal2015b}. The quantities $\mbh$ and $\sigma$ for NGC 4486B were taken from \cite{Sagliaetal2016}. These values are different from those given by \cite{KormendyHo2013}; we fitted the data using both values,
resulting in small differences. Besides these, we add eight more galaxies from \cite{Sagliaetal2016} (S+; see table \ref{tab:Galaxytabel}).}}}}

{{{{We do not consider galaxies for which the black hole mass was derived from the width of the H$\alpha$ emission line rather than dynamical models. (This is the case for  RGG~118 (LEDA 87300), which is interesting by itself. See, e.g., \citealt{Grahametal2016b}.)}}}}

\begin{figure*}
\begin{center}
{\includegraphics{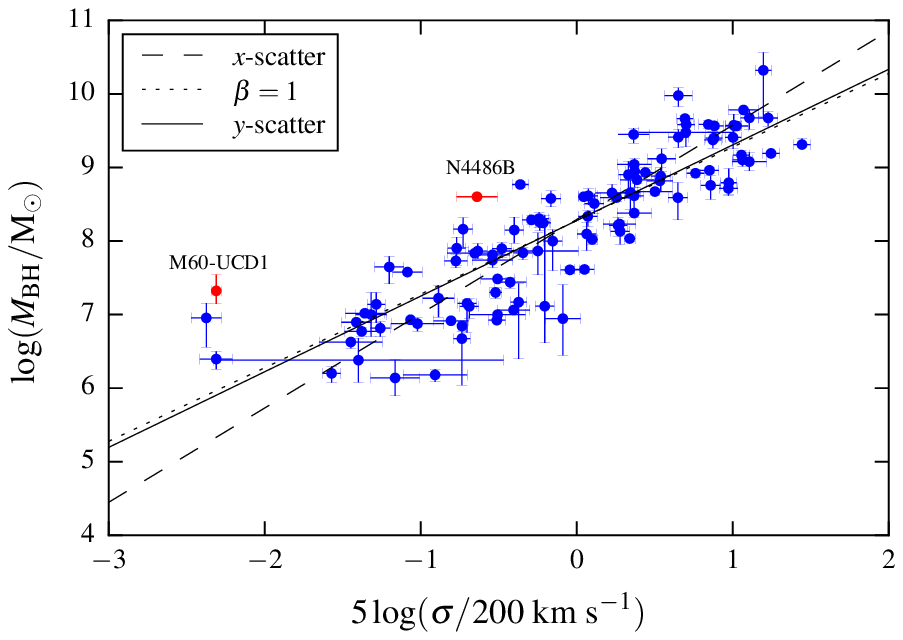}}\\[-0.15cm]
{\includegraphics{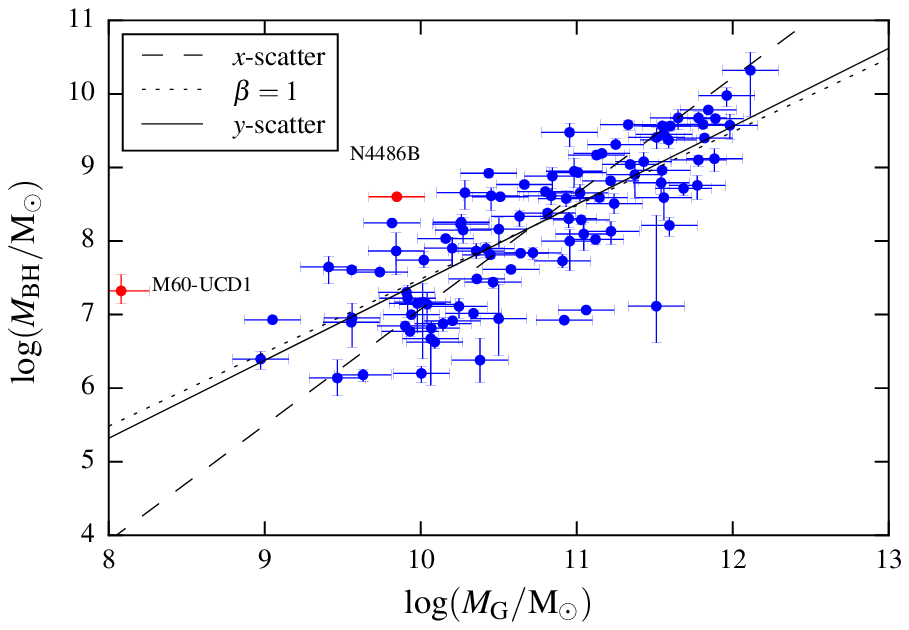}}\\[-0.15cm]
\end{center}
\vspace{-0.85cm}
\caption{The $M_{\rm BH}$--$\sigma^5$ (top; {{{{106}}}} galaxies) and the $M_{\rm BH}$--$M_{\rm G}$ (bottom {{{{103}}}} galaxies) relations for our sample. $M_{\rm BH}$, $M_{\rm G}$, and $\sigma$ are the black hole mass, bulge mass, and stellar velocity dispersion of the galaxies, respectively. See text for the sources of objects.
The dashed and solid lines are the best-fitting linear relations (in log-log scale) using $x$-and $y$-scatter directions respectively {{{{(see \citealt{Lahavetal2011} for this technical point).}}}} The dotted line has a slope of $\beta=1$. The logarithm is base 10.}
\label{fig:corsigmamg}
\end{figure*}

Many studies established that the black hole mass correlates strongly with the stellar velocity dispersion, i.e., the $M_{\rm BH}$--$\sigma$ correlation, and with the bulge mass, i.e., the $M_{\rm BH}$--$M_{\rm G}$ correlation (e.g. \citealt{KormendyRichstone1995, Magorrian_etal1998,  Gebhardt_etal2000, MerrittFerrarese2001, Tremaine_etal2002, Wandel2002, Laor2001, Hu2009, Gultekin_etal2009, Graham2009, Graham2011, Greene_etal2010,  McConnellMa2013, KormendyHo2013, Savorgnanetal2016}). In Fig. \ref{fig:corsigmamg} we present these two correlations for our sample.
The {{{{103}}}} galaxies in the $M_{\rm BH}$--$M_{\rm G}$ plot (Fig. \ref{fig:corsigmamg}, bottom panel) are only those which had a measurement of their bulge mass, while the $M_{\rm BH}$--$\sigma$ plot (Fig. \ref{fig:corsigmamg}, top panel) contains the entire sample of {{{{106}}}} galaxies.

We also calculated the best-fitting parameters for each correlation in three ways, depending on the scatter, as explained by  \cite{Lahavetal2011} and \cite{SokerMeiron2011}. We obtained results which are similar to those obtained in recent years (e.g. \citealt{SokerMeiron2011, Graham2011, McConnellMa2013, Lahavetal2011, Savorgnanetal2016}), and hence will not present them here.

\section{THE SCATTER}
\label{sec:scatternew}

The (total) `scatter' of the data points is quantified by the root mean square of the residuals from the line of best fit, while the intrinsic scatter is a measure of the natural spread of the data. The total scatter is not identical to the intrinsic scatter. Hence, the intrinsic scatter contributes to the total scatter, but so do measurement errors, thus making the latter larger. Usually the intrinsic scatter is estimated from the logarithmic data, and hence is dimensionless, but below we refer to the scatter (extrinsic and intrinsic) in the linear scale, {{{{which isn't.}}}}

Let us now proceed to examine the proposal that the $M_{\rm BH}$--$M_{\rm G}$ correlation is the result of mergers. We take the fundamental building-blocks to be $n$ galaxies with identically and independently distributed bulge masses, around a mean $M_{{\rm G},0}$. The initial mass distribution of the black holes within these galaxies has an average $M_{{\rm BH},0}$ with variance $s_0^2$. After
$n$ mergers, the mean bulge mass becomes $M_{\rm G}=n M_{{\rm G},0}$. According to the central limit theorem, the masses of the black
holes are normally distributed around $M_{\rm BH}=n M_{{\rm BH},0}$ with
variance $s^2 = n s_0^2$, for any given number of merges $n \gg 1$ (or a given $M_{\rm G}$), whatever the initial distribution. In this simple model, the relation between black hole and
bulge mass is indeed linear -- $M_{\rm BH} = M_{\rm G} (M_{{\rm BH},0} / M_{{\rm G},0})$. More importantly for our analysis, the intrinsic scatter increases as $\sqrt{M_{\rm G}}$.
This is consistent with the results of \citet{Hirschmann}, who investigated the intrinsic scatter using cosmological halo merger trees \citep{Genel}.

According to the merger scenario $M_{\rm BH} \propto M_{\rm G}$, and hence
$\sqrt{M_{\rm BH}} \propto \sqrt{M_{\rm G}}$, too. We use error propagation to calculate the (intrinsic) scatter $\varepsilon_0$ of $\sqrt{M_{\rm BH}}$ for a given $\sqrt{M_{\rm G}}$
\begin{equation}
\varepsilon_0 =
\left( \frac {{\rm d}\sqrt{M_{\rm BH}}}{{\rm d} M_{\rm BH}} \right) s
= \frac{s}{2 \sqrt{M_{\rm BH}}}
= \frac{\sqrt{n} s_0}{2 \sqrt{n M_{{\rm BH},0}}}
= \frac{s_0}{2 \sqrt{M_{{\rm BH},0}}}
= {\rm const.}
\label{eq:Delta1}
\end{equation}
Therefore, this merger-based scenario for the co-evolution of the galaxy and its SMBH
predicts an $M_{\rm BH}^{1/2}$--$M_{\rm G}^{1/2}$ relation with a scatter that
does not depend on the bulge mass.

\begin{figure*}
\hspace{-0.4cm}
\centering
\includegraphics[width=235pt]{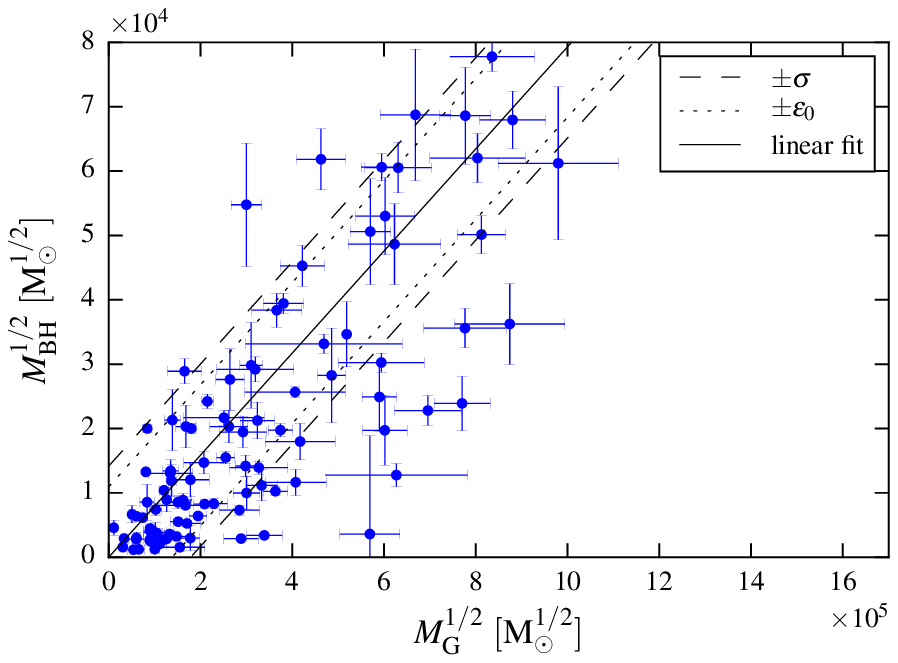}
\includegraphics[width=235pt]{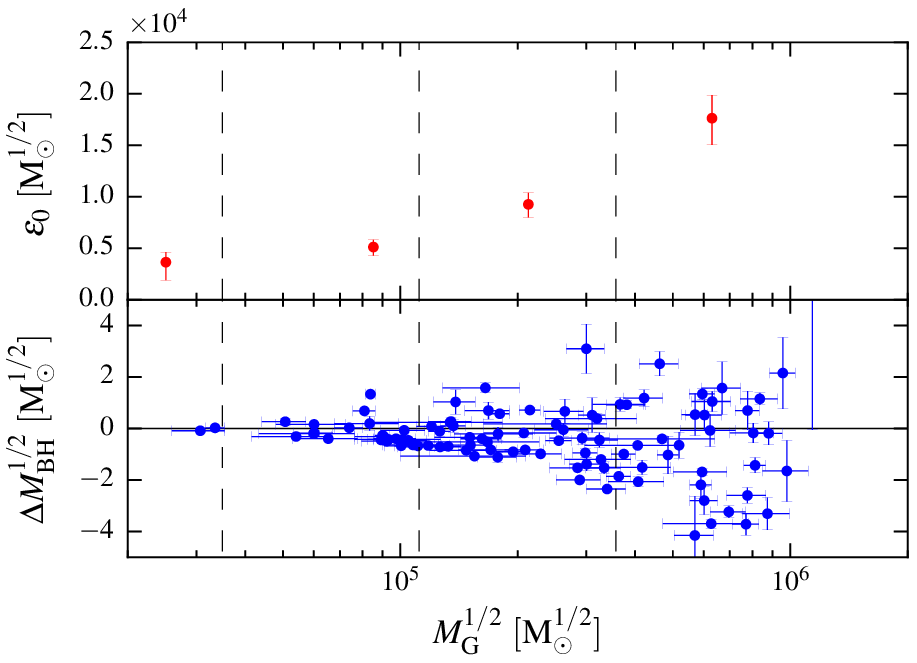}
\caption{Left: A plot of $M_{\rm BH}^{1/2}$ versus $M_{\rm G}^{1/2}$,
where $M_{\rm BH}$ and  $M_{\rm G}$ are the SMBH and host galaxy bulge masses,
respectively. The solid line represents the linear fit, the dashed lines are one standard deviation contours, whereas the dotted lines represent the intrinsic scatter of all data points. Right: the bottom panel shows the vertical distance of the data points from the best-fitting line versus $M_{\rm G}^{1/2}$; the upper panel shows the intrinsic scatter for each of the four bins. These bins contain 3, 23, 42 and 35 objects, beginning at low masses.}
\label{fig:cor2}
\end{figure*}

We test this merger-based scenario and its consequence of a constant scatter of $M_{\rm BH}^{1/2}$ by {{{{examining the $M_{\rm BH}^{1/2}$--$M_{\rm G}^{1/2}$ correlation (left panel of Fig. \ref{fig:cor2})}}}}.
The data {{{{are}}}} the same as in Fig. \ref{fig:corsigmamg}, but on a linear scale rather than logarithmic. The solid line is the best linear fit (zero intercept).

In the lower part of the right panel of Fig. \ref{fig:cor2}, we plot the residuals; and evidently the total scatter around the fit-line increases with bulge mass. In order to investigate the intrinsic scatter's dependence on mass, we divided the data into four equal-logarithmic-width bins of $M_{\rm G}$. For each bin, we calculated the intrinsic scatter in $M_{\rm BH}$ required to bring the reduced sum of square residuals (from the ridge line derived from the entire data-set) to 1. We calculate the errors on $\varepsilon_0$ from the shape of the $\chi^2$ distribution \citep{SokerMeiron2011}.
The upper part of the right panel of Fig. \ref{fig:cor2} shows that the intrinsic scatter also increases with $\mgal$, contrary to what is expected in a merger-only scenario.

There is room for a scenario where the correlation is formed by mergers alone, while separate hypothetical processes lead to the observed scatter but do not contribute to the correlation. We examined this possibility by generating mock samples with statistical properties similar to the galaxy sample; the BH masses were put on the correlation line and randomly shifted in such a way that the scatter in the $M_{\rm BH}^{1/2}$--$M_{\rm G}^{1/2}$ plane was constant and equal to $5\times 10^3\,\mathrm{M}_\odot^{1/2}$, corresponding to the leftmost bin in the right panel Fig. \ref{fig:cor2} (the premise being that if those hypothetical processes are turned off, the intrinsic scatter would be smaller than or equal to the value at the low-mass end). We then tried to reproduce the behaviour of the scatter in Fig. \ref{fig:cor2} by additionally scattering the data points in various reasonable ways (i.e. BH mass can only increase), imitating these hypothetical mass-dependent processes. Our finding was that this normally changes the correlation so significantly, that a straight line in the $M_{\rm BH}^{1/2}$--$M_{\rm G}^{1/2}$ plane is not a good fit any more, and the residuals show a clear trend. This is only circumvented if there are two very finely tuned processes that separately modify the BH and bulge masses in such a way that the linear correlation is preserved. It is far more likely that the observed scatter is a feature of the process that forms the correlation, rather than two additional finely tuned processes.
\section{OUTLIERS WITH MISSING STELLAR MASS}
\label{sec:outliers}
\subsection{The outliers}
\label{subsec:motiviation}

Our conclusion from the analysis presented in Section \ref{sec:scatternew} is that mergers play a small role in establishing the correlation between the BH masses and the stellar masses of their host galaxies bulges \citep[in agreement with][]{Lahavetal2011}. The alternative is an AGN feedback mechanism.
The AGN, mostly through jets, can heat up the gas (e.g., \citealt{ZubovasKing2012, DiazSantosetal2016}), as occurred for example in cooling flow clusters,
and expel the gas, as is expected during galaxy formation (e.g., \citealt{Boweretal2008}).
Not all of the gas is expelled during galaxy formation, though -- some of the inter-stellar medium (ISM) that is heated by AGN activity stays bound, but expands. The expanded ISM suffers radiative cooling and flows back to the galaxy. One outcome is that a cooling flow can take place at galaxy formation \citep{Soker2010}.

The expanded gas has a lower density and resides in a smaller depth of the potential well relative to its initial state. Hence, it is more vulnerable to external perturbations as it is moving through the intra-cluster medium (ICM). We speculate that in rare cases, a young galaxy might pass very close to the central galaxy of its cluster, or near another very massive galaxy, where the ICM is relatively dense, and undergo ram-pressure stripping, {{{{which should be more efficient there. (We assume that the SMBH acquires a substantial amount of its mass early in the galactic history. See, e.g. \citealt{FerreMateuetal2015})}}}}.
ISM removal should inhibit star formation at later times; the outcome of which is a galaxy with a normal BH mass, but with less than normal stellar mass. The galaxy would be on the $M_{\rm BH}$--$\sigma$ correlation because the velocity dispersion is determined mainly by the dark matter. However, {{{{it}}}} would be deficient in stellar mass, and hence might be an outlier {{{{in}}}} the $M_{\rm BH}$--$M_{\rm G}$ correlation -- an over-massive BH galaxy.
This scenario of ISM inflation, followed by ram-pressure stripping, is different from that proposed by \cite{Fabianetal2013}, who did not mention stripping, but rather the inability to gather material due to the galaxy's rapid motion in the cluster core.

{{{{Two noteworthy over-massive outliers in the $M_{\rm BH}$--$M_{\rm G}$ plot (Fig. \ref{fig:corsigmamg}) are NGC 4486B and M60-UCD1. In projection, NGC 4486B resides near the giant elliptical galaxy M~87 at the centre of the Virgo Cluster -- less than $40 \kpc$ from the centre. Although it has been suggested that it had been tidally stripped (e.g. \citealt{Faber1973,Barberetal2016}), this case serves as part of our motivation. Recent measurements by \cite{Sagliaetal2016} place it as an outlier in both $\mbh$-$\mgal$ and $\mbh$-$\sigma$ correlations; whereas the differing data in \cite{KormendyHo2013} imply that it is only an outlier} {{{in}}} {the $\mbh$-$\mgal$ correlation -- as would be the case if it had undergone the process we suggest. On the other hand, the galaxy M60-UCD1 is as extreme an outlier as NGC 4486B in the $M_{\rm BH}$--$M_{\rm G}$ correlation, but it is also far from the $\mbh$-$\sigma$ ridge line, hence it could have suffered mainly tidal stripping, as suggested by \cite{Sethetal2014}. Further motivation for the proposed scenario is discussed next. }}}}

\begin{figure}
\begin{center}
{{{\includegraphics{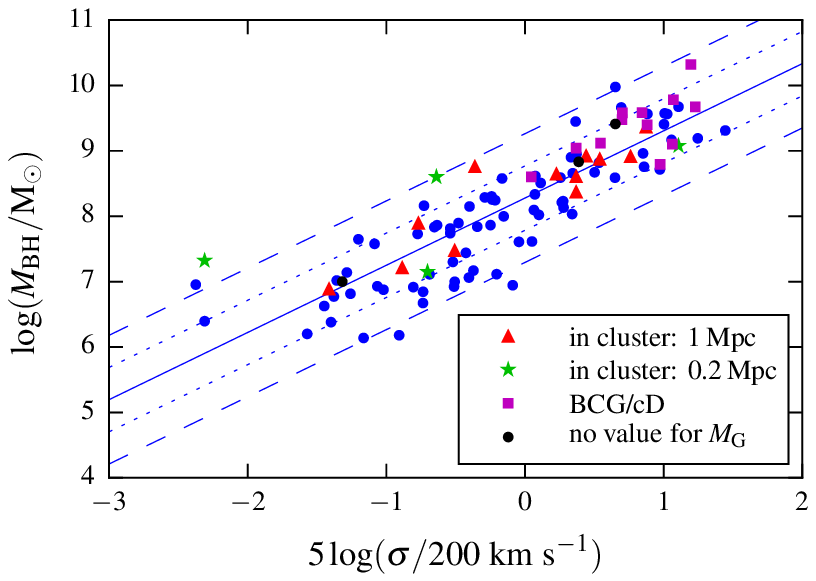}}}}
\includegraphics{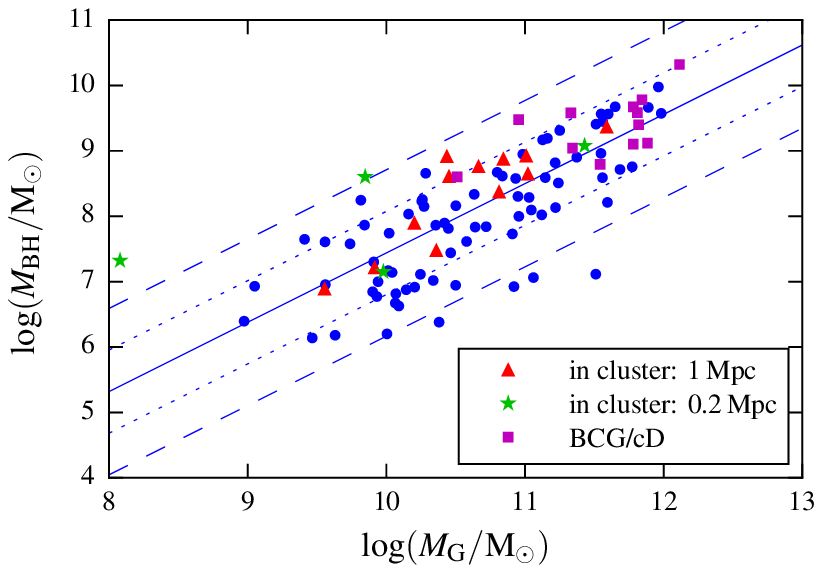}
\end{center}
\caption{ {{{{ Top:
The $\mbh$-$\sigma$ correlation (same as the upper panel of Fig. \ref{fig:corsigmamg}), with galaxies in cluster indicated as follows: central galaxies (magenta squares), galaxies within projected 0.2\,Mpc (green stars) and 1\,Mpc (red triangles) from their cluster's centre. Other galaxies are indicated by blue circles. For clarity, error bars are not shown (but see Fig. \ref{fig:corsigmamg}). Best fitting line ($y$-scatter) as well as 1- and 2-standard deviation contours are shown in solid, dotted and dashed lines, respectively. }}}}
Bottom: the same for the $\mbh$-$\mgal$ correlation, for the same galaxies.
Note that cluster galaxies are systematically above the trend line in the $\mbh$-$\mgal$ correlation, but not in the $\mbh$-$\sigma$ correlation.
}
\label{fig:cluster}
\end{figure}

We further classify the galaxies that are cluster members by their projected distance from the centre of their host cluster. In Fig. \ref{fig:cluster} we mark those with projected distance less than $1 \Mpc$, those with projected distance less than $0.2 \Mpc$, and the central cluster galaxies.
For cluster member classification, we used the \emph{HyperLeda} database \citep{Makarovetal2014}\footnote{http://leda.univ-lyon1.fr/}. Then we calculated the distance between the galaxy and its cluster's centre using co-ordinates from \emph{SIMBAD} and \emph{VizieR}\footnote{http://simbad.u-strasbg.fr/simbad/, http://vizier.u-strasbg.fr/} \citep{Wengereral2000,Ochsenbeinetal2000}. The distances to these galaxies were taken from \cite{KormendyHo2013} and \cite{Savorgnanetal2016}, with preference to the former in case of discrepancies. While the sample is very small, we find that galaxies that reside near their host cluster's centre tend to occupy the upper parts of the {{{{$\mbh$-$\mgal$ plot (bottom panel of Fig. \ref{fig:cluster}), i.e., above the trend line, in agreement with \cite{McGee2013}}}}}. We interpret this as a deficiency of stellar mass. For Virgo Cluster galaxies, we used the minimum distance from M 87, M 60 or M 49 (the latter two are $\sim 6 \times 10^{11} M_\odot$). Although M 87 is the central galaxy, we consider the two others as also capable harbouring a dense ICM environment.
{{{{Unlike the case of the $\mbh$-$\mgal$ correlation, in the $\mbh$-$\sigma$ correlation (upper panel of Fig. \ref{fig:cluster}) we do not discern such a trend, i.e., the cluster galaxies are symmetrically distributed around the correlation line in the $\mbh$-$\sigma$. This is in contrast with the findings of \cite{ZubovasKing2012} and \cite{McGee2013}, perhaps due to a difference in samples and the small numbers of cluster galaxies. We infer therefore that galaxies near the centres of clusters show a stronger trend of stellar mass deficiency than of having too low velocity dispersions. This might hint that ram-pressure stripping, which only removes gas, is more significant than tidal stripping in accounting for this trend, if at all.}}}}

\subsection{The stripping scenario}
\label{subsec:scenario}

{{{{We can estimate the order of magnitude of some of the parameters of the ram-pressure scenario, which exemplify the important parts of the scenario and their respective roles. Tidal stripping would also be important, but we emphasise the additional role of the efficient ram-pressure stripping as a result of the inflation of the ISM by jets. However, to solidify the proposed scenario, simulations of cluster evolution would have to be conducted.}}}}

Many galaxies in clusters have elongated orbits. Galaxies spend only a fraction of the time near the centre, where they can be efficiently stripped of their gas. The inflated phase of the gas in the galaxy takes about the free fall time from the radius outside of which most of the gas is to be stripped. We found this radius to be of the order of $27 \kpc$, for the density profiles we use (see below). The fall-back time of the inflated envelope -- when the inner half-mass of the ISM returns to its original radius -- is $\tau_b \approx 10^8 \yr$. A galaxy with an elongated orbit to about $0.5 \Mpc$, would take about 5 times as long to reach the cluster centre. Over all, only galaxies residing near the cluster centre (within about $0.5 \Mpc$) would undergo stripping -- only if their orbit {{{{carried}}}} them further in -- and even then the stripping phase might last for only part of the orbit. Based on these numbers, we expect only about {{{{a}}}} fifth, say $\sim 10-30 \%$, of the galaxies within $\sim 0.5 \Mpc$ to be stellar-deficient outliers.

{{{{We assume that this holds true at the period of major AGN activity, at a redshift of $z \approx 2$. Although the cluster, or smaller proto-cluster, was probably not fully virialized at that time, galaxy velocities were not much different from their values at present. In some cases the stripping process took place in a small group that later became the centre of the cluster -- that already contained a massive elliptical galaxy at its centre. The subsequent analysis is therefore valid only for galaxies for which these conditions hold. There is evidence that some clusters were already formed by $z \approx 3$ and even earlier (e.g., \citealt{Wylezaleketal2014}), as well as observations of a proto-cluster at $z= 3.786$ \citep{Deyetal2016}.  By $z \approx 2$ a hot ICM gas could have been created in these clusters, e.g., as in the cluster JKCS~041 at redshift $z=1.803$ \citep{Andreonetal2014}, and the cores of the clusters could have hosted a characteristic population of galaxies \citep{Strazzulloetal2015}. Concerning NGC 4486B, it is possible that the Virgo cluster in which it resides was partially formed, and that it had already contained about $10\%$ of its mass \citep{Sorceetal2016}.
Generally, our ram-pressure stripping scenario overlaps with the period during which a feedback mechanism determines the correlation between the SMBH mass and the galactic properties.  }}}}

{{{{Our analysis below does not depend much on the redshift in the relevant range, from the period of major AGN activity at $z \approx 2$ to the present, for the following reasons. First, as noted above, the velocities of galaxies in the cluster are about the same in the redshift range $z=0-2$. Secondly, the density in the inner regions of the cluster did not change much with respect to physical radius. Although the average density within the virial radius decreases strongly with time, the density of the dark matter, and to some degree the density of the baryonic matter, did not change with respect to physical radius, both in the cluster and in the stripped galaxy. That is, the dark matter within the radius of $r \la 0.5 \Mpc$, and the baryonic matter density in the region where most stripping takes place in the cluster, $r \la 0.2 \Mpc$, did not change much from $z \approx 2$ to present.  As we are interested in a stripping near the cD galaxy (or a massive elliptical galaxy), the process could take place at the high redshift of $z \approx 2$ when the virial radius of the proto-cluster was much smaller than its value at present, but proto-clusters already existed. In short, the differences of the relevant parameters with redshift are much smaller than the uncertainties in the proposed scenario that we check here with a toy model.}}}}

To demonstrate the feasibility of our ram-pressure stripping idea, we put forward a very simple toy model. Consider a galaxy with a spherical density profile given by $\rho (r) = \rho_{\rm g}(r) + \rhor(r)$, where $\rho_{\rm g}$ is the gas density and $\rhor$ is the density of all other matter in the galaxy. We further assume that AGN activity inflates the ISM by a uniform factor $\lambda > 1$, while keeping the total gaseous mass the same. The new gas density profile is
\begin{equation}\label{eq:densexp}
\rhog'(r) = \lambda^{-3} \rho_\mathrm{g}(r/\lambda).
\end{equation}
For the stripping efficiency we use the analytical fit derived by \cite{McCarthyetal2008} from their numerical simulations.
According to \cite{McCarthyetal2008} the radius beyond which gas is removed by ram-pressure stripping, $R_\textrm{strip}$, is given by
\begin{equation}\label{eq:rst}
  P_\textrm{ram} = \alpha \frac{GM(R_\textrm{strip})\rhog(R_\textrm{strip})}{R_\textrm{strip}},
\end{equation}
where $\alpha$ is a constant which they find to be roughly 2. The ram-pressure that is exerted by the ICM is defined by $P_\textrm{ram} = \rho_{\rm ICM} v^2$, where $\rho_{\rm ICM}$ is the ICM density and $v$ is the speed of the galaxy through the ICM.
We take for the ICM density $\rho_{\rm ICM} = 2 \times 10^{-28} \g \cm^{-3}$, corresponding to the electron number density $n_e=1 \times 10^{-4} \cm^{-3}$ at a radius of $500 \kpc$ in the Virgo cluster \citep{Urbanetal2011}. 

The ISM in the galaxy to be stripped is assumed to be distributed according to a $\beta$-model \citep{Roedigeretal2015}
\begin{equation}\label{egn:gas_beta_model}
  \rhog(r) = \frac{\rho_{0\textrm{,g}}}{[1+(r/r_0)^2]^{3/4}},
\end{equation}
with $r_0 = 2 \kpc$ and a central density of $\rho_{0\textrm{,g}} =  2 \times 10^{-25} \g \cm^{-3}$ (typical values). For the total baryonic mass, which at this early phase is mainly gas, we take $2 \times 10^{10} M_\odot$. From the density profile, total gas mass, and the central density used above we find that the ISM extends up to radius of $R_{\rm ISM}=40 \kpc$. We take the dark matter to be distributed according to a Navarro--Frenk--White (NFW) profile \citep{Navarroetal1997}
\begin{equation}\label{eqn:mass}
  \rho_\mathrm{DM}(r) = \frac{\rho_0}{(r/a)(1+r/a)^2}.
\end{equation}
For our model galaxy we use a total dark matter mass $M_{\rm DM} = 10^{11} M_\odot$, and (like \citealt{McCarthyetal2008}) take the cut-off radius to be $r_{200} = c_{200} a$ the radius at which the dark matter density is 200 times the critical density; $c_{200}$ is calculated from $M_\mathrm{DM}$ using the relation given by \cite{Netoetal2007}. Once it is determined, $\rho_0$ can be calculated from the definition of $r_{200}$; then $a$ is deduced from the total mass and $\rho_0$. This yields $\rho_0 = 2.5 \times 10^{-24} {\g \cm^{-3}}, a = 5.2 {\kpc}, r_{200} = 54.3 \kpc$.

We take the ISM gas and gravitational potential according to these parameters, then let the gas expand by a factor $\lambda$ according to equation \eqref{eq:densexp}, and use equation \eqref{eq:rst} with a galaxy velocity through the ICM of $v = 700 \km \s^{-1}$ to calculate how much mass is stripped.
We emphasise that we use these typical parameters to demonstrate the feasibility of the proposed ram-pressure scenario. We are not in the stage of performing full scale numerical simulations of the scenario {{{{that tracks the evolution with redshift}}}} -- this is a task for a future study.

We now present our findings of the fraction $f(\lambda)$ of the ISM gas that is retained after ram-pressure stripping, to the total gaseous mass before both stripping and ISM inflation, defined in the following way:
\begin{equation}
f(\lambda) = \frac{M'_\mathrm{g}(\rst')}{M_\mathrm{g}}.
\end{equation}
We find that for the parameters we used, before any expansion of the ISM, the ram-stripping by the ICM removes gas outside of a radius of $27 \kpc$, and leaves a fraction of $f(\lambda=1)= 0.46$ from the ISM. If before the galaxy is stripped the AGN inflates the ISM by a factor of $\lambda = 1.2$, $1.4$, and $1.6$, the stripping takes place from radii of $22 \kpc$, $19 \kpc$, and $16 \kpc$, respectively, and the fraction of gaseous mass that is not stripped off is only $f(\lambda=1.2)= 0.26$, $f(\lambda=1.4)= 0.16$, and $f(\lambda=1.6)= 0.1$, respectively. {{{{The exact values might change somewhat when the evolution of the cluster with redshift is considered. However, as discussed above, the changes are smaller than the uncertainties in the toy model.}}}}

In the above calculation we considered only inflated gas. However, a considerable portion of the ISM might be ejected by the AGN activity itself, without any need for the ram-pressure stripping {{{{\citep{ZubovasKing2012}}}}}. For the present goals, we have demonstrated that ram pressure stripping could in principle account for outliers in the $M_{\rm BH}$--$M_{\rm G}$ correlation that are not outliers in the $M_{\rm BH}$--$\sigma$ correlation.

We now estimate the amount of energy that is required to inflate the ISM.
If the gas is inflated, its potential energy increases; this energy must come from the SMBH. According to the virial theorem $\frac{1}{2} \Delta U = \Delta E$ and only the change in potential energy needs be calculated for an order of magnitude estimate. We assume also that the potential $\varphi$ is mainly due to the dark matter, and that it is given as an NFW potential. Thus, the energy of the gas is
\begin{equation}\label{eqn:energy}
  U \simeq \int\rhog \varphi \mathrm{d}^3 \mathbf{r},
\end{equation}
where $\rhog$ is only the gas density. The change in the energy of the inflated ISM is given then by
\begin{equation}\label{eqn:energy_change}
  \Delta U \simeq \int [\rhog'(r) - \rhog(r)]\varphi(r) \cdot 4\pi r^2 \mathrm{d}r.
\end{equation}
We find that for the parameters of our model, to expand the ISM by a factor of two in size requires an energy of $\Delta E \approx 1.2 \times 10^{57} \erg$. If this energy were to come from the AGN alone, then with an efficiency of $1\%$, the total mass accreted by the black hole would be $\Delta M \approx 6 \times 10^4 M_\odot$. As discussed above, the expansion factor can be less than two.

\section{Summary}
\label{sec:summary}

Having examined the claim that mergers of low mass galaxies are the main process behind the SMBH mass to bulge mass correlation, we conclude that if it causes the $M_{\rm BH}$--$M_{\rm G}$ correlation, then it also creates an intrinsic scatter in the
$M_{\rm BH}^{1/2}$--$M_{\rm G}^{1/2}$ relation as given by equation (\ref{eq:Delta1}) -- a constant scatter in this plane.
To test this expectation, we used a sample of {{{{103}}}} galaxies that was compiled from recent studies (see Table \ref{tab:Galaxytabel}), and found that the intrinsic scatter in $M_{\rm BH}^{1/2}$--$M_{\rm G}^{1/2}$ increases faster than it should have, had it been driven mainy by mergers (see Fig. \ref{fig:cor2}). As all galaxies participate in mergers, we included all kinds of galaxies in our study. \cite{Lahavetal2011} performed a similar analysis for different types of galaxies separately and found that the results hold for each subgroup separately. We do not dispute the claim that mergers influence the various correlations to some degree. We argue though, that in view of the conclusions of Section \ref{sec:scatternew}, mergers cannot be the dominant processes behind the correlations.

We also considered another process that might play a secondary role: we looked at galaxies that tend to have missing stellar mass -- namely, they have small ratios of bulge mass to SMBH mass, but are not far from the correlation of the SMBH mass with the velocity dispersion. In Fig. \ref{fig:cluster}, we classified cluster members according to their distance from the centre of their cluster, and saw that those very close to it (or to another very large elliptical galaxy in the cluster) tend to occupy the upper parts of the $\mbh$-$\mgal$ correlation. We proposed that ram-pressure stripping plays a part in forming these outliers (Section \ref{subsec:motiviation}) -- it removes mainly the gas and hence primarily influences the stellar mass.
The stripping becomes significant when AGN activity inflates the gas, thus making it more vulnerable, in conjunction with the galaxy's moving in the dense central region of a cluster. Only a small number of galaxies are expected to be strongly affected by this process. {{{{It has been noticed before that the environment can influence the correlations studied here. \cite{McGee2013}, for example, found that the slopes of the correlations depend on whether the galaxy is close to the centre of a cluster, or it is far from the centre, a satellite galaxy.
\cite{McGee2013} also mentioned that ram-pressure stripping can affect Satellite galaxies in-falling into the ICM,  but also cited \cite{ZubovasKing2012} in that stripping affects mainly the outskirts of the ISM. We are mostly concerned with gas that would have stayed inside the galaxy, were it not inflated by the AGN and stripped off right afterwards. This is an additional effect to the one described in these papers. The inflation by the AGN jets may enhance the difference between cluster galaxies and others.}}}}

In cooling flows, the main role of the AGN is to heat the ICM, and in galaxy formation it is to eject the gas. We speculated here that inflation of the ISM is another important process by which the AGN determines the properties of its host galaxy. This speculation is motivated by our finding that mergers play only a small role in determining the correlations, and by the properties of galaxies with small distances to cluster centres.

\medskip

{{{{ We thank the anonymous referee for very helpful comments.}}}}
This research was supported by the Asher Fund for Space Research at the Technion, and in part by the European Research Council under the European Union's Horizon 2020 Programme, ERC-2014-STG grant GalNUC 638435. YM acknowledges support from the China Postdoctoral Science Foundation through grant No. \textit{2015T80011} and from the Strategic Priority Research Program ``The Emergence of Cosmological Structure'' of the Chinese Academy of Sciences (No. \textit{XDB09000000}) (Pilot B programme).

\onecolumn

\begin{deluxetable}{lrrrrrll}
\tablecolumns{8}
\small
\tablewidth{0pt}
\tabletypesize{\scriptsize}
\tablecaption{Our sample of 106 galaxies\label{tab:Galaxytabel}}
\tablehead{
\colhead{Galaxy} &
\colhead{$\sigma$} &
\colhead{$M_{\rm BH}$-high} &
\colhead{$M_{\rm BH}$-low} &
\colhead{$M_{\rm BH}$} &
\colhead{$M_{\rm G}$} &
\colhead{Cluster} &
\colhead{Reference}
\\
\colhead{} &
\colhead{[\kms]} &
\colhead{[$10^8 M_\odot$]} &
\colhead{[$10^8 M_\odot$]} &
\colhead{[$10^8 M_\odot$]} &
\colhead{[$10^{10} M_\odot$]} &
\colhead{} &
\colhead{}
}
\startdata
\astrobj{
A1836BCG
}&$ 295 \pm 8 $&$ 43 $&$ 34 $&$ 38 $&$ 65 \pm 20 $& A $1836_{BCG}$ & KH, SG, MM \\
\astrobj{
A3565BCG
}&$ 326 \pm 9 $&$ 15 $&$ 10 $&$ 13 $&$ 60 \pm 10 $& A $3565_{BCG}$ & KH, SG, MM \\
\astrobj{
Circinus
}&$ 132 \pm 7 $&$ 0.018 $&$ 0.012 $&$ 0.015 $&$ 0.43 \pm 0.2 $& & KH, SG, MM \\
\astrobj{
IC 1459
}&$ 317 \pm 7 $&$ 35 $&$ 18 $&$ 26 $&$ 33 \pm 5 $& & KH, SG, MM \\
\astrobj{
IC 2560
}&$ 143 \pm 4 $&$ 0.073 $&$ 0.011 $&$ 0.047 $&$ 1.2 \pm 0.4 $& & KH, SG \\
\astrobj{
M 31
}&$ 166 \pm 5 $&$ 2.1 $&$ 0.94 $&$ 1.4 $&$ 1.9 \pm 0.3 $& & KH, SG, MM \\
\astrobj{
M 32
}&$ 69 \pm 2 $&$ 0.032 $&$ 0.018 $&$ 0.025 $&$ 0.094 \pm 0.03 $& & KH, SG, MM \\
\astrobj{
M 81
}&$ 149 \pm 4 $&$ 0.92 $&$ 0.58 $&$ 0.73 $&$ 2.3 \pm 0.9 $& & KH, SG, MM \\
\astrobj{
M 87
}&$ 327 \pm 14 $&$ 64 $&$ 57 $&$ 61 $&$ 70 \pm 20 $& Virgo$_{cD}$ & KH, SG, MM \\
\astrobj{
Milky Way
}&$ 103 \pm 10 $&$ 0.044 $&$ 0.035 $&$ 0.042 $&$ 1.2 \pm 0.3 $& & KH, SG, MM \\
\astrobj{
NGC 0253
}&$ 109 \pm 5 $&$ 0.2 $&$ 0.05 $&$ 0.1 $ & & & SG\\
\astrobj{
NGC 0524
}&$ 245 \pm 7 $&$ 9.9 $&$ 7.6 $&$ 8.5 $&$ 10 \pm 5 $& UGCl 24$_{1}$ & KH, SG, MM \\
\astrobj{
NGC 0821
}&$ 206 \pm 6 $&$ 1.8 $&$ 0.74 $&$ 1.2 $&$ 11 \pm 2 $& & KH, SG, MM \\
\astrobj{
NGC 1023
}&$ 205 \pm 6 $&$ 0.45 $&$ 0.37 $&$ 0.41 $&$ 3.8 \pm 0.8 $& & KH, SG, MM \\
\astrobj{
NGC 1068
}&$ 158 \pm 4 $&$ 0.088 $&$ 0.08 $&$ 0.084 $&$ 8.3 \pm 2 $& & KH, SG \\
\astrobj{
NGC 1194
}&$ 148 \pm 12 $&$ 0.71 $&$ 0.65 $&$ 0.68 $&$ 4.4 \pm 1 $& UGCl 49 & KH, SG, MM \\
\astrobj{
NGC 1277
}&$ 333 \pm 17 $&$ 65 $&$ 33 $&$ 49 $&$ 26.9 $& Perseus$_{1,2}$ & KH ($\sigma$ only) \\
\astrobj{
NGC 1300
}&$ 178 \pm 5 $&$ 1.3 $&$ 0.35 $&$ 0.73 $&$ 0.7 \pm 0.3 $& & KH, SG, MM \\
\astrobj{
NGC 1316
}&$ 226 \pm 6 $&$ 2.1 $&$ 1.2 $&$ 1.6 $&$ 39 \pm 20 $& Fornax & KH, SG, MM \\
\astrobj{
NGC 1332
}&$ 325 \pm 8 $&$ 17 $&$ 13 $&$ 15 $&$ 13 \pm 4 $& & KH, SG, MM \\
\astrobj{
NGC 1374
}&$ 169 \pm 4 $&$ 6.4 $&$ 5.3 $&$ 5.9 $&$ 4.6 \pm 0.6 $& Fornax$_{1}$ & KH, SG, MM \\
\astrobj{
NGC 1399
}&$ 313 \pm 7 $&$ 9.6 $&$ 4.3 $&$ 6.2 $&$ 35 \pm 4 $& Fornax$_{BCG}$ & KH, SG, MM \\
\astrobj{
NGC 1407
}&$ 275 \pm 7 $&$ 52 $&$ 39 $&$ 46 $&$ 77 \pm 10 $& & KH, SG, MM \\
\astrobj{
NGC 1550
}&$ 276 \pm 7 $&$ 44 $&$ 32 $&$ 38 $&$ 21 \pm 5 $& UCGl 56$_{BCG}$ & KH, SG, MM \\
\astrobj{
NGC 2273
}&$ 138 \pm 7 $&$ 0.087 $&$ 0.078 $&$ 0.082 $&$ 1.6 \pm 0.7 $& & KH, SG, MM \\
\astrobj{
NGC 2549
}&$ 145 \pm 4 $&$ 0.16 $&$ 0.057 $&$ 0.14 $&$ 0.95 \pm 0.2 $& UGCl 116$_{1,2}$ & KH, SG, MM \\
\astrobj{
NGC 2748
}&$ 115 \pm 5 $&$ 0.62 $&$ 0.26 $&$ 0.44 $&$ 0.26 \pm 0.07 $& & KH \\
\astrobj{
NGC 2778
}&$ 169 \pm 4 $&$ 0.27 $&$ 0.025 $&$ 0.15 $&$ 1 \pm 0.5 $& & KH, SG \\
\astrobj{
NGC 2787
}&$ 196 \pm 5 $&$ 0.45 $&$ 0.36 $&$ 0.41 $&$ 0.36 \pm 0.1 $& & KH, SG, MM \\
\astrobj{
NGC 2960
}&$ 166 \pm 8 $&$ 0.12 $&$ 0.11 $&$ 0.12 $&$ 11 \pm 3 $& & KH, SG, MM \\
\astrobj{
NGC 2974
}&$ 227 \pm 11 $&$ 1.9 $&$ 1.5 $&$ 1.7 $&$ 1.8 \pm 1 $& & SG \\
\astrobj{
NGC 3079
}&$ 105 \pm 45 $&$ 0.048 $&$ 0.012 $&$ 0.024 $&$ 2.4 \pm 2 $& UGCl 163& SG \\
\astrobj{
NGC 3091
}&$ 300 \pm 8 $&$ 39 $&$ 34 $&$ 37 $&$ 35 \pm 5 $& & KH, SG, MM \\
\astrobj{
NGC 3115
}&$ 237 \pm 7 $&$ 13 $&$ 5.4 $&$ 8.9 $&$ 9.6 \pm 2 $& & KH, SG, MM \\
\astrobj{
NGC 3227
}&$ 133 \pm 6 $&$ 0.25 $&$ 0.093 $&$ 0.17 $&$ 0.82 \pm 0.3 $& UGCl 169$_{1}$ & KH, SG, MM \\
\astrobj{
NGC 3245
}&$ 207 \pm 6 $&$ 2.6 $&$ 1.6 $&$ 2.2 $&$ 4.3 \pm 2 $& & KH, SG, MM \\
\astrobj{
NGC 3368
}&$ 125 \pm 10 $&$ 0.09 $&$ 0.06 $&$ 0.075 $&$ 1.4 \pm 0.2 $& & KH, SG, MM \\
\astrobj{
NGC 3377
}&$ 143 \pm 4 $&$ 2.1 $&$ 0.82 $&$ 1.5 $&$ 3.2 \pm 1 $& & KH, SG, MM \\
\astrobj{
NGC 3379
}&$ 207 \pm 6 $&$ 5.2 $&$ 3.1 $&$ 4.1 $&$ 6.9 \pm 2 $& & KH, SG, MM \\
\astrobj{
NGC 3384
}&$ 146 \pm 4 $&$ 0.17 $&$ 0.09 $&$ 0.13 $&$ 1.8 \pm 0.3 $& & KH, SG, MM \\
\astrobj{
NGC 3393
}&$ 164 \pm 5 $&$ 0.32 $&$ 0.23 $&$ 0.28 $&$ 2.9 \pm 1 $& & KH, SG, MM \\
\astrobj{
NGC 3414
}&$ 237 \pm 12 $&$ 2.7 $&$ 2.1 $&$ 2.4 $&$ 6.5 \pm 1 $& UGCl 192$_{1}$ & SG \\
\astrobj{
NGC 3489
}&$ 106 \pm 5 $&$ 0.068 $&$ 0.051 $&$ 0.059 $&$ 0.85 \pm 0.2 $& & KH, SG, MM \\
\astrobj{
NGC 3585
}&$ 211 \pm 6 $&$ 4.4 $&$ 2.3 $&$ 3.2 $&$ 17 \pm 6 $& & KH, SG, MM \\
\astrobj{
NGC 3607
}&$ 227 \pm 6 $&$ 1.8 $&$ 0.9 $&$ 1.4 $&$ 17 \pm 5 $& & KH, SG, MM \\
\astrobj{
NGC 3608
}&$ 185 \pm 5 $&$ 4.8 $&$ 2.9 $&$ 3.8 $&$ 8.6 \pm 3 $& & KH, SG, MM \\
\astrobj{
NGC 3842
}&$ 270 \pm 11 $&$ 120 $&$ 67 $&$ 95 $&$ 92 \pm 10 $& UGCl 234 & KH, SG, MM \\
\astrobj{
NGC 3945
}&$ 192 \pm 10 $&$ 0.26 $&$ 0 $&$ 0.088 $&$ 3.2 \pm 0.7 $& UGCl 229 & KH \\
\astrobj{
NGC 3998
}&$ 284 \pm 7 $&$ 9.5 $&$ 7.3 $&$ 8.3 $&$ 2.7 \pm 1 $& UGCl 229$_{1}$ & KH, SG, MM \\
\astrobj{
NGC 4026
}&$ 179 \pm 5 $&$ 2.3 $&$ 1.4 $&$ 1.8 $&$ 1.8 \pm 0.5 $& UGCl 229 & KH, SG, MM \\
\astrobj{
NGC 4151
}&$ 156 \pm 8 $&$ 0.72 $&$ 0.58 $&$ 0.65 $&$ 2.8 \pm 2 $& & SG \\
\astrobj{
NGC 4258
}&$ 121 \pm 5 $&$ 0.38 $&$ 0.37 $&$ 0.38 $&$ 0.55 \pm 0.09 $& & KH, SG, MM \\
\astrobj{
NGC 4261
}&$ 313 \pm 9 $&$ 6.3 $&$ 4.1 $&$ 5.2 $&$ 48 \pm 10 $& Virgo & KH, SG, MM \\
\astrobj{
NGC 4291
}&$ 256 \pm 7 $&$ 10 $&$ 4.7 $&$ 7.6 $&$ 7 \pm 2 $& UGCl 269$_{1}$ & KH, SG, MM \\
\astrobj{
NGC 4342
}&$ 234 \pm 7 $&$ 6.7 $&$ 2.7 $&$ 4.5 $&$ 1.9 \pm 0.5 $& Virgo & KH, SG, MM \\
\astrobj{
NGC 4374
}&$ 296 \pm 8 $&$ 10 $&$ 8.3 $&$ 9.2 $&$ 35 \pm 10 $& Virgo & KH, SG, MM \\
\astrobj{
NGC 4382
}&$ 182 \pm 5 $&$ 2.2 $&$ 0 $&$ 0.13 $&$ 32 \pm 7 $& Virgo & KH \\
\astrobj{
NGC 4388
}&$ 104 \pm 5 $&$ 0.081 $&$ 0.077 $&$ 0.079 $&$ 0.36 \pm 0.2 $& Virgo$_{1}$ & KH, SG, MM \\
\astrobj{
NGC 4459
}&$ 171 \pm 5 $&$ 0.83 $&$ 0.56 $&$ 0.69 $&$ 5.2 \pm 1 $& Virgo & KH, SG, MM \\
\astrobj{
NGC 4472
}&$ 300 \pm 7 $&$ 28 $&$ 22 $&$ 25 $&$ 66 \pm 9 $& Virgo$_{cD}$ & KH, SG, MM \\
\astrobj{
NGC 4473
}&$ 186 \pm 5 $&$ 1.4 $&$ 0.4 $&$ 1 $&$ 9 \pm 2 $& Virgo & KH, SG, MM \\
\astrobj{
NGC 4486A
}&$ 111 \pm 3 $&$ 0.2 $&$ 0.077 $&$ 0.14 $&$ 1.1 \pm 0.3 $& Virgo & KH, SG, MM \\
\astrobj{
NGC 4486B
}&$ 149 \pm 9 $&$ 4.2 $&$ 3.8 $&$ 4 $&$ 0.7 \pm 0.04 $& Virgo$_{1,2}$ & S+ \\
\astrobj{
NGC 4526
}&$ 222 \pm 11 $&$ 5.9 $&$ 3.5 $&$ 4.5 $&$ 10 \pm 2 $& Virgo$_{1}$ & KH \\
\astrobj{
NGC 4552
}&$ 252 \pm 13 $&$ 5.2 $&$ 4.2 $&$ 4.7 $&$ 6.3 \pm 4 $& Virgo & SG \\
\astrobj{
NGC 4564
}&$ 160 \pm 5 $&$ 0.96 $&$ 0.6 $&$ 0.79 $&$ 2.6 \pm 0.6 $& Virgo & KH, SG, MM \\
\astrobj{
NGC 4594
}&$ 256 \pm 8 $&$ 7 $&$ 6.1 $&$ 6.6 $&$ 16 \pm 9 $& & KH, SG, MM \\
\astrobj{
NGC 4596
}&$ 140 \pm 4 $&$ 1.1 $&$ 0.46 $&$ 0.8 $&$ 1.6 \pm 0.2 $& Virgo$_{1}$ & KH, SG, MM \\
\astrobj{
NGC 4621
}&$ 225 \pm 11 $&$ 4.3 $&$ 3.5 $&$ 3.9 $&$ 14 \pm 2 $& Virgo & SG \\
\astrobj{
NGC 4649
}&$ 352 \pm 10 $&$ 57 $&$ 37 $&$ 47 $&$ 60 \pm 9 $& Virgo$_{cD}$ & KH, SG, MM \\
\astrobj{
NGC 4697
}&$ 175 \pm 5 $&$ 2.2 $&$ 1.7 $&$ 1.9 $&$ 11 \pm 4 $& & KH, SG, MM \\
\astrobj{
NGC 4736
}&$ 112 \pm 3 $&$ 0.08 $&$ 0.05 $&$ 0.065 $&$ 1.2 \pm 0.2 $& & KH, SG, MM \\
\astrobj{
NGC 4751
}&$ 355 \pm 9 $&$ 17 $&$ 14 $&$ 16 $&$ 14 \pm 3 $& & KH, SG \\
\astrobj{
NGC 4826
}&$ 97 \pm 3 $&$ 0.02 $&$ 0.012 $&$ 0.016 $&$ 1 \pm 0.2 $& & KH, SG, MM \\
\astrobj{
NGC 4889
}&$ 347 \pm 8 $&$ 370 $&$ 50 $&$ 210 $&$ 130 \pm 20 $& Coma$_{BCG}$ & KH, SG, MM\\
\astrobj{
NGC 4945
}&$ 117 \pm 3 $&$ 0.024 $&$ 0.0078 $&$ 0.014 $&$ 0.29 \pm 0.1 $& & KH, SG \\
\astrobj{
NGC 5077
}&$ 233 \pm 7 $&$ 12 $&$ 3.8 $&$ 8 $&$ 24 \pm 3 $& & KH, SG, MM \\
\astrobj{
NGC 5128
}&$ 140 \pm 4 $&$ 0.66 $&$ 0.43 $&$ 0.54 $&$ 8.1 \pm 2 $& & KH, SG, MM \\
\astrobj{
NGC 5328
}&$ 333 \pm 9 $&$ 61 $&$ 33 $&$ 47 $&$ 45 \pm 10 $& & KH, SG \\
\astrobj{
NGC 5516
}&$ 321 \pm 11 $&$ 42 $&$ 32 $&$ 37 $&$ 40 \pm 9 $& & KH, SG, MM \\
\astrobj{
NGC 5576
}&$ 179 \pm 5 $&$ 2.5 $&$ 1.5 $&$ 2 $&$ 8.9 \pm 1 $& & KH, SG, MM \\
\astrobj{
NGC 5813
}&$ 239 \pm 12 $&$ 7.5 $&$ 6.1 $&$ 6.8 $ & & & SG\\
\astrobj{
NGC 5845
}&$ 237 \pm 7 $&$ 5.3 $&$ 2.6 $&$ 4.1 $&$ 2.8 \pm 0.7 $& UGCl 352$_{1}$ & KH, SG, MM \\
\astrobj{
NGC 5846
}&$ 237 \pm 12 $&$ 12 $&$ 10 $&$ 11 $&$ 22 \pm 20 $& UGCl 352$_{BCG}$ & SG \\
\astrobj{
NGC 6086
}&$ 318 \pm 8 $&$ 53 $&$ 24 $&$ 37 $&$ 96 \pm 30 $& & KH, SG, MM \\
\astrobj{
NGC 6251
}&$ 297 \pm 8 $&$ 7.7 $&$ 3.7 $&$ 5.7 $&$ 59 \pm 9 $& UGCl 388 & KH, SG, MM \\
\astrobj{
NGC 6264
}&$ 158 \pm 8 $&$ 0.31 $&$ 0.3 $&$ 0.31 $&$ 2.3 \pm 0.5 $& AWM 5$_{1}$ & KH, SG, MM \\
\astrobj{
NGC 6323
}&$ 158 \pm 13 $&$ 0.1 $&$ 0.099 $&$ 0.1 $&$ 0.87 \pm 0.2 $& UGCl 420 & KH, SG, MM \\
\astrobj{
NGC 6861
}&$ 389 \pm 10 $&$ 25 $&$ 19 $&$ 21 $&$ 18 \pm 4 $& & KH, SG \\
\astrobj{
NGC 7052
}&$ 270 \pm 8 $&$ 6.2 $&$ 1.9 $&$ 3.9 $&$ 36 \pm 6 $& & KH, SG, MM \\
\astrobj{
NGC 7457
}&$ 67 \pm 3 $&$ 0.14 $&$ 0.036 $&$ 0.09 $&$ 0.36 \pm 0.08 $& & KH \\
\astrobj{
NGC 7582
}&$ 156 \pm 9 $&$ 0.72 $&$ 0.44 $&$ 0.55 $&$ 1 \pm 0.3 $& & KH, SG, MM \\
\astrobj{
NGC 7619
}&$ 299 \pm 7 $&$ 31 $&$ 18 $&$ 24 $&$ 39 \pm 10 $& UGCl 487$_{1}$ & KH, SG, MM \\
\astrobj{
NGC 7768
}&$ 257 \pm 11 $&$ 18 $&$ 8.8 $&$ 13 $&$ 76 \pm 20 $& UGCl 493$_{BCG}$ & KH, SG, MM \\
\astrobj{
UGC 3789
}&$ 107 \pm 6 $&$ 0.11 $&$ 0.095 $&$ 0.1 $&$ 2.2 \pm 0.8 $& UGCl 100 & KH, SG, MM \\
\astrobj{
Cygnus A
}&$ 270 \pm 8 $&$ 33 $&$ 19 $&$ 26 $ & & & KH, SG\\
\astrobj{
NGC 1271
}&$ 276 \pm 39 $&$ 40 $&$ 19 $&$ 30 $ & $9 \pm 2$ & Perseus$_{1,2}$ & \\
\astrobj{
M60-UCD1
}&$ 69 \pm 1 $&$ 0.35 $&$ 0.14 $&$ 0.21 $&$ 0.012 \pm 0.003 $& Virgo$_{1,2}$ &  \\
\astrobj{
NGC 307
}&$ 204 \pm 3 $&$ 4.3 $&$ 3.7 $&$ 4 $&$ 3.2 \pm 0.4 $& UGCl 14$_{1}$ & S+ \\
\astrobj{
NGC 1398
}&$ 234 \pm 4 $&$ 1.2 $&$ 0.98 $&$ 1.1 $&$ 1.4 \pm 0.2 $& & S+ \\
\astrobj{
NGC 3627
}&$ 122 \pm 1 $&$ 0.087 $&$ 0.083 $&$ 0.085 $&$ 0.11 \pm 0.01 $& & S+ \\
\astrobj{
NGC 3923
}&$ 237 \pm 11 $&$ 34 $&$ 22 $&$ 28 $&$ 36 \pm 8 $& & S+ \\
\astrobj{
NGC 4371
}&$ 143 \pm 2 $&$ 0.08 $&$ 0.06 $&$ 0.07 $&$ 0.8 \pm 0.07 $& & S+ \\
\astrobj{
NGC 4501
}&$ 157 \pm 3 $&$ 0.23 $&$ 0.17 $&$ 0.2 $&$ 0.81 \pm 0.07 $& & S+ \\
\astrobj{
NGC 4699
}&$ 181 \pm 4 $&$ 1.9 $&$ 1.7 $&$ 1.8 $&$ 0.65 \pm 0.09 $& & S+ \\
\astrobj{
NGC 5018
}&$ 209 \pm 3 $&$ 1.2 $&$ 0.91 $&$ 1.1 $&$ 13 \pm 2 $& & S+ \\

\enddata
\tablenotetext{{}}{
$M_{\rm BH}$-high and -low are the upper and lower limits of
the SMBH mass, respectively, according to the data available in the references as detailed
in Section \ref{sec:samplenew}. The second column on the right gives the bulge mass,
which is the stellar mass of the spheroidal component. {{{{In the seventh column one could see if the galaxy is a member of a cluster (to our knowledge); subscripts indicate if it is a BCG/cD, if it is $1 \Mpc$ (1) or $200 \kpc$ (2) from the centre. Those with no subscripts are in clusters but too far away from the centre. See Section \ref{subsec:motiviation}. }}}} }
\end{deluxetable}

\end{document}